\documentclass[twocolumn]{IEEEtran}
\usepackage{amsmath}
\usepackage{amsthm}
\usepackage{amsfonts}
\usepackage{amssymb}
\usepackage{caption}
\usepackage{mathrsfs}
\usepackage{cases}
\usepackage{cite}
\usepackage{graphicx}
\usepackage{mathrsfs}
\usepackage{subfigure}
\usepackage{epstopdf}
\usepackage{color}

\usepackage{enumitem}
\usepackage{setspace}
\usepackage{bm}

\theoremstyle{remark}

\newtheorem{remark}{\hspace{1em}Remark}

\begin{document}

\title{\LARGE  Identifying the Fake Base Station: A Location Based Approach
\thanks{
The work was supported in part by the
National Natural Science Foundation of China under Grant 61671364 and in part
by the Outstanding Young Research Fund of Shaanxi Province.
}
\thanks{
K.-W. Huang and H.-M. Wang are with the Ministry
of Education Key Lab for Intelligent Networks and Network Security, Xi'an Jiaotong
University, Xi'an, 710049, Shaanxi, P. R. China. Email: {\tt
xjtu-huangkw@outlook.com,
xjbswhm@gmail.com}.
}
\author{Ke-Wen Huang, \hspace{0.05in} Hui-Ming Wang,~\IEEEmembership{Senior Member,~IEEE}
}
}
\maketitle

\begin{abstract}
    Fake base station (FBS) attack is a great security challenge to wireless user equipment (UE). During the cell selection stage, the UE receives multiple synchronization signals (SSs) from multiple nearby base stations (BSs), and then synchronizes itself with the strongest SS. A FBS also can transmit a SS with sufficient power to confuse the UE, which makes the UE connect to the FBS, and may lead to the leakage of private information.
    In this letter, countermeasure to the FBS attack by utilizing the location information is investigated.  Two location awareness based FBS-resistance schemes are proposed by checking the received signal strength according to the position of the UE and a legitimate BS map.
    The successful cheating rate (SCR) definded as the probability that the UE will connect to the FBS is investigated.
    Numeric results show that with the two proposed schemes, the SCR can be greatly reduced especially when the transmit power of the FBS is large. Beyond that,
    a cooperation aided method is further proposed to improve the performance, and
    we show that the cooperation aided method can further suppress the SCR when the signal strength from the FBS is similar to that from the legitimate BS.
\end{abstract}
\begin{IEEEkeywords}
Fake base station, location awareness, physical layer authentication, detection.
\end{IEEEkeywords}

\section{Introduction}
\label{Sec:Introduction}
The initial cell selection (CS) stage during which an user equipment (UE) searches for a suitable base station (BS) to camp on is vulnerable to a fake base station (FBS) attack.
In the CS stage, the UE keeps listening to the wireless broadcast channel and searches for the synchronizing signal (SS) from surrounding base stations (BSs).
After that, the UE selects a suitable BS based on the received SSs, and then begins to establish a wireless connection with the BS \cite{Book:S.Sesia}.
If a FBS transmits a spoofing SS during the CS stage with sufficiently high power ({referred to as the SS spoofing attack \cite{M.LichtmanCM2016}}), the UE may be attracted by and attempt to camp on the FBS rather than any legitimate BS (LBS) \cite{M.Labib2015,M.Labib2016}.
Currently, the widely-used authentication method for the UE to distinguish the FBS from the LBSs is based on encryption.
However, though the FBS will fail during the key-based authentication procedure, it still significantly degrades the system performance, for example, the FBS can lead to the significant increase of the delay before the UE can successfully connect to the legitimate network and more seriously, if the FBS further spoof the control signal, it can even cause the access denial to the network \cite{M.Labib2015,M.Labib2016}.

Recently, using the physical layer parameters to authenticate the signal source, namely physical layer authentication, has gained considerable attention which does not depends on any private key, see \cite{X.Wang2016} and references therein \footnote{We note that physical layer based approach has also been extensive investigated to secure data transmission in wireless channel, namely physical layer security, as a complementary scheme to the conventional cryptography-based scheme, see \cite{YLiu2017CST} and references therein.}.
For example, channel impulse response and device fingerprint based schemes were proposed in  \cite{L.Xiao2008} and \cite{A.C.Polak2011}, respectively, to verify the transmitter's identity.
However, these methods are almost impossible to be exploited to combat with the SS spoofing attack because they require the pre-recorded estimations of the physical layer parameters which is impossible to be obtained  during the CS stage, as in this stage, there is exactly no wireless connection between the UE and any LBS.
To combat with the SS spoofing attack, in this letter, we propose a location-based physical layer approach.
To the best of our knowledge, no existing literature is specifically focused on this issue.

Our proposed schemes require that the UE knows its own position and the locations of the LBSs. Note that this requirement is not hard to satisfy in view of the facts that nowadays, the UE (such as a smartphone) is usually equipped with a positioning module (GPS) so that the UE can know its own position, and
the locations of the LBSs are generally fixed and invariant which can be also known by the UE via an off-line map (GoogleMap in a smartphone).

Our main idea is that with the location information of the UE and LBSs, taking the path loss, the shadowing effect and the small-scale fading into consideration, the average received synchronization signal strength (ARSSS) from LBSs should be within a proper range. On the other hand, a FBS usually transmits with a high power level to attract the UE \cite{M.LichtmanCM2016}. The UE can check the ARSSS and
once the ARSSS exceed its normal range, it is reasonable to suspect that the corresponding SS is transmitted from a FBS. Note that location awareness has been proposed to improve the performance of future networks, e.g., in \cite{M.Koivisto}, but has not been utilized for the UEs to identify the FBS.
The major benefit of the proposed methods in this letter is that once the ARSSSs are obtained, they can make a decision immediately, while the conventional key-based methods require the UE to synchronize itself with the SS possibly transmitted by the FBS.
The contributions of this letter are summarized as follows:
1) based on the location information, we provide two different ARSSS checking criterions, which are referred as suspicious ARSSS region (SAR) criterion  and maximum-likelihood (ML) criterion, respectively;
2) by taking the shadow fading and small-scale fading effects into consideration, the successful cheating rates (SCR), defined as the probability that a given UE will connect to the FBS, are derived for our proposed methods;
3) a cooperative ARSSS checking scheme is proposed to improve the performance when the ARSSS from the FBS approaches to that from the closest LBS.

\section{System Model and Problem Statement}
In this section, we first introduce our system model. Then we evaluate the SCR of a given UE when there is no security mechanism during the CS stage.

\begin{figure}
    \centering
    \includegraphics[width=1.7 in, height = 0.8 in]{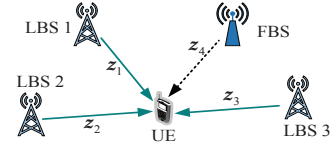}
    \caption{\small System model with three LBSs and one FBS.}
    \label{Model}
    \vspace{-6mm}
\end{figure}

A comprehensive system model is given in Fig. \ref{Model}.
We consider that there are $M$ LBSs, referred to as LBS $1$, LBS $2$, $\cdots$, LBS $M$, and a FBS around the considered UE broadcasting mutually orthgonal SSs, denoted by $\mathcal{Z}\triangleq\{\bm{z}_1,\bm{z}_2,\cdots,\bm{z}_{M+1}\}$, to enable the nearby UEs to connect to them. Assume $\bm{z}_{m}\in\mathcal{C}^{\tau}$, $m\in\{1,2,\cdots,M\}$, and $\bm{z}_{M+1}\in\mathcal{C}^{\tau}$ are the SS transmitted by the $m$-th LBS and the FBS, where $\tau$ is the length of the SSs satisfying $\tau\geq M+1$.
For $m=1,2,\cdots,M+1$, we assume that $\|\bm{z}_m\|^2=\tau$.
During the CS stage, the UE first searches the existence of the SSs and obtains the set $\mathcal{Z}$ \footnote{This can be simply realized by matching all the possible SSs with the received signal and check whether the output power exceeds a pre-designed threshold. For simplicity, we assume all the SSs in $\mathcal{Z}$ can be successfully detected in this letter.}.
Then it keeps listening to the SSs for a total of $L$ observing time slots (TSs) and record the ARSSS of each SS. Finally, we assume the UE synchronize itself to the SS with largest ARSSS
\footnote{The 3GPP specification requires the UE to search for the strongest cell except for the some special cases such as when the strongest cell is in ``barred'' status \cite{3GPP}. For simplicity, in this letter, we only consider the situation where the UE is able to connect to the strongest cell to provide a basic understanding of the effect of the FBS attack. Other special cases are more complicated and are left for future research.}.

{
The received signal at the UE during the $l$-th TS is
\begin{align}
\bm{y}\left[l\right]&=\sum_{m=1}^{M+1}\sqrt{P_m\Psi_{m}}d_{m}^{-\alpha/2}h_m\left[l\right]\bm{z}_m +\bm{n}\left[l\right],\label{Simultaneousarrive}
\end{align}
where $\{d_m,h_m[l],\Psi_m, P_m\}$, for $m\in\{1,2,\cdots,M\}$, and $\{d_{M+1},h_{M+1}[l],\Psi_{M+1},P_{M+1}\}$ are the distance, the Rayleigh small scale fading factor during the $l$-th TS, the log-normal shadow fading, and the transmit power of the $m$-th LBS and the FBS,  respectively, $\bm{n}\left[l\right]\sim \mathcal{CN}\left(\bm{0},\sigma^2\bm{I}\right)$ is the received noise, and $\alpha$ is the exponential factor of the path loss.
We assume the log-normal shadow fading factors remain unchanged during the whole $L$ TSs, and they are identically and independently distributed as $\mathcal{N}\left(0, \sigma_\Psi^2\right)$ in decibels, and all the Rayleigh small scale fading factors are identically and independently distributed as $\mathcal{CN}\left(0, \sigma_h^2\right)$.
For simplicity, we assume that the LBSs transmit with the same power $P$, i.e., $P=P_1=P_2=\cdots=P_M$.

Matching $\bm{y}\left[l\right]$ with $\bm{z}_m$, for $m\in\{1,2,\cdots,M+1\}$, the UE obtains the output power, which can be written as
\begin{align}
p_{m}\left[l\right]&\triangleq |\bm{y}\left[l\right]\bm{z}_m^H|^2/\tau= \left|\sqrt{P_m\Psi_{m}d_{m}^{-\alpha}}h_m\left[l\right]+n_m\left[l\right]\right|^2,\nonumber
\end{align}
where $n_m\left[l\right]\triangleq\bm{n}\left[l\right]\bm{z}_m^H/\tau^2\sim\mathcal{CN}\left(0,\sigma^2/\tau\right)$.
{Note that in general, the SSs are designed to be detectable at a low SNR \cite{M.LichtmanCM2016},} and therefore for simplicity, we assume that the noise term $n_m\left[l\right]$ can be neglected.
In fact, by increasing $\tau$, the impact of the noise gets smaller.
Therefore, we have $p_{m}\left[l\right]\approx P_md_{m}^{-\alpha}\Psi_{m}\left|h_m\left[l\right]\right|^2$, for $m\in\{1,2,\cdots,M+1\}$.

The ARSSS of the $m$-th SS, i.e., $\bm{z}_m$, is defined as
\begin{align}
\mathrm{S}_m &\triangleq \frac{1}{L}\sum_{l=1}^{L}\mathrm{lt}\left(p_{m}\left[l\right]\right)=
U_m + \frac{1}{L}\sum_{l=1}^{L}X_m\left[l\right],\label{ARSSSExpression}
\end{align}
where we have $\mathrm{lt}\left(x\right)\triangleq10\log_{10}\left(x\right)$,
$X_m\left[l\right]\triangleq 2\mathrm{lt}\left(\left|h_m\left[l\right]\right|\right)$, and $U_m\triangleq\mathrm{lt}\left(P_m\right) - \alpha\mathrm{lt}\left( d_m\right)+\mathrm{lt}\left( \Psi_m\right)$.
}

We have to emphasize here that at this stage, the UE only obtain the set of SSs from surrounding BSs (both LBSs and FBS), i.e., $\mathcal{Z}$, and the corresponding set of the ARSSSs, denoted by $\mathcal{S}\triangleq\left\{\mathrm{S}_1,\mathrm{S}_2,\cdots,\mathrm{S}_{M+1}\right\}$, but for each received SS, the UE does not know which BS is the signal source.

Based on \eqref{ARSSSExpression}, we can evaluate the SCR when there is no authentication mechanism for the UE to distinguish the FBS from the LBSs during the CS stage.
We denote the SCR as $\mathcal{P}_{S}$, then we have
\begin{align}
\nonumber
\mathcal{P}_{S}&\triangleq\mathcal{P}\left\{\mathrm{S}_{M+1}>\hat{S}\right\} =\int_{-\infty}^{+\infty}f_{M+1}\left(x\right)\prod_{m=1}^MF_m\left(x\right)\mathrm{d}x,
\end{align}
where $\hat{S} \triangleq \max\limits_{1\leq m\leq M}\mathrm{S}_m$, and for $1\leq m\leq M+1$, $f_{m}(x)$ and $F_{m}(x)$ are the probability density function (PDF) and cumulative distribution function (CDF) of $S_m$, respectively.

According to the central-limit theorem, as $L$ becomes large, $S_m$ is asymptotically distributed as $\mathcal{N}\left(u_m,\sigma_S^2\right)$, where $u_m\triangleq \mathrm{lt} (\sigma_h^2P_m) - \alpha\mathrm{lt} \left(d_m\right) - \gamma$ and $\sigma_S^2\triangleq \sigma_\Psi^2 + \frac{1}{L}\sigma_X^2$ with $\gamma$ being the Eular's constant and $\sigma_X^2\triangleq \left(\gamma + \mathrm{lt}\left(\sigma_h^2\right)\right)^2 + \frac{\pi^2}{6}$. Therefore, we have
\begin{align}
\mathcal{P}_{S}\doteq
\frac{1}{\sqrt{2\pi}}\int_{-\infty}^{+\infty}e^{-\frac{t^2}{2}}\prod_{m=1}^M
\Phi\left(t+\frac{u_{M+1}-u_m}{\sigma_S}\right)\mathrm{d}x,\nonumber
\end{align}
where $\Phi(x)$ is the CDF of a standard Gaussian random variable, and ``$\doteq$'' means ``asymptotically equals to'' under the condition that $L\rightarrow+\infty$.
It should be pointed out that, if $u_{M+1}-u_{m}$ is sufficiently large for $m=1,2,\cdots,M$, then $\mathcal{P}_{S}\rightarrow 1$, which means that the UE almost always choose to connect to the FBS.
\begin{remark}
{ In practice, the different SSs from the different BSs usually arrive at the UE in an asynchronous manner.
However, for simplicity of mathematical description, we assumed that the different SSs simultaneously arrive the UE as in \eqref{Simultaneousarrive}.
Note that this simplification does not change the basic process during the CS stage.}
\qed
\end{remark}

\section{Identify the FBS Through ARSSS}
In this section, based on the UE's own position and the prior locations information of the LBSs according to an off-line map, we provide two practical methods to combat with the SS spoofing attack at the UE-side during the CS stage.
With the location information, we assume the UE knows $u_m$ for $m=1,2,\cdots,M$ and $\sigma_S^2$.

Once the UE obtains $\mathcal{Z}$ and $\mathcal{S}$, we provide the following two criterions which allow the UE to choose a SS from the observed SS set $\mathcal{Z}$ in a more secure manner. The basic idea is that by checking the relative locations of the UE itself and LBSs in the map, the ARSSS from  LBSs should within a proper range. A significant large ARSSS could be suspected to be from the FBS.
For notational simplicity, in this section, we denote the final SS chosen by the UE as $\bm{z}_{*}$.

\subsection{SAR based ARSSS checking}
As in current cellular networks, the UE always searches the strongest cell, the greedy FBS may transmit with a higher power level than the LBSs to attract the UE \cite{M.LichtmanCM2016}.
Therefore, we define a SAR for the ARSSSs in $\mathcal{S}$, which is denoted by $\mathcal{I}\triangleq\left(\bar{S},+\infty\right)$.
More specifically,
all the elements in $\mathcal{S}$ that are within $\mathcal{I}$ will be suspected to be from the FBS,
and the UE should synchronize itself to the strongest SS in $\mathcal{S}\setminus\mathcal{I}$.
Here, $\bar{S}$ is chosen such that
$\mathcal{P}\left\{ \hat{S} > \bar{S} \right\}\leq\delta$ with $0<\delta<1$ being a small pre-designed value, which is similar to the false alarm rate in hypothesis test theory \cite{M.Barkat}.
Note that we have
$\mathcal{P}\left\{\hat{S}>\bar{S}\right\}\doteq1 - \prod_{m=1}^M \left(1- Q\left(\frac{\bar{S}-u_m}{\sigma_S}\right)\right)$, and the value of $\bar{S}$ can be searched through a bisection method. As a result, in the SAR based method, we have $\bm{z}_{*} = \bm{z}_{\hat{m}}$ with $S_{\hat{m}} =
\mathrm{argmax}_{s\in\mathcal{S}\setminus\mathcal{I}}~s$.

As we have assumed that $\bm{z}_{M+1}$ is transmitted by the FBS, under the proposed strategy, the SCR satisfies
\begin{align}
&\mathcal{P}_{S}=\mathcal{P}\left\{\bm{z}_{*}=\bm{z}_{M+1}\right\}\overset{(a)}{\leq}
\mathcal{P}\left\{\hat{S}<S_{M+1}< \bar{S}\right\} +\delta\nonumber \\
&\doteq
\frac{1}{\sqrt{2\pi}}\int_{-\infty}^{\frac{\bar{S}-u_{M+1}}{\sigma_S}}e^{-\frac{t^2}{2}}
\prod_{m=1}^M\Phi\left(t+\frac{u_{M+1}-u_m}{\sigma_S}\right)\mathrm{d}t  +\delta,\nonumber
\end{align}
where the inequality in $(a)$ is obtained by assuming that the UE will always connect to the FBS if $\hat{S}>\bar{S}$.
In the following two reasonable scenarios, $\bar{S}$ can be approximated by some simple and computationally efficient expressions:
\subsubsection{when the UE is much closer to one of the LBS}
For example, when the UE is much closer to LBS 1, then we have $u_1\gg u_m$ for $m=2,3,\cdots,M$, and $\hat{S}\approx S_1$. Therefore, we have $\mathcal{P}\left\{\hat{S}>\bar{S}\right\}\approx \mathcal{P}\left\{S_1>\bar{S}\right\}\doteq Q\left(\frac{\bar{S}-u_1}{\sigma_S}\right)$. Accordingly, we can obtain $\bar{S}\approx \sigma_SQ^{-1}\left(\delta\right)+u_1$.

\subsubsection{when the UE is at the coverage edge of several LBSs}
For example, when the UE is at the coverage edge of LBS 1, LBS 2,$\cdots$, and LBS K, then
we have $u_1\approx u_2\approx\cdots\approx u_K\gg u_{j}$, for $K+1\leq j\leq M$. Therefore, we have $\mathcal{P}\left\{\hat{S}>\bar{S}\right\}\approx \mathcal{P}\left\{\max\limits_{1\leq k\leq K} S_k>\bar{S}\right\}\approx 1 - \left(1 - Q\left(\frac{\bar{S}-u_1}{\sigma_S}\right)\right)^K$. Accordingly, we can obtain $\bar{S}\approx \sigma_SQ^{-1}\left(1 - \sqrt[k]{1-\delta}\right)+u_1$.

\subsection{ML based ARSSS checking}
In this subsection, we introduce a ML based method for the UE to choose a proper SS.
Different from the SAR based method where the ARSSSs are compared with a pre-designed threshold, the
ML based method determines a SS by directly checking the likelihood function \cite{M.Barkat}.
More specifically, $\bm{z}_{*}$ satisfies
$\bm{z}_{*} = \bm{z}_{\check{m}}$ with $\check{m}=\mathrm{argmax}_{m}~f_{\mathrm{max}}\left(S_{m}\right)$ and $f_{\mathrm{max}}\left(x\right)=\sum_{k=1}^M f_k\left(x\right) \prod_{m=1,m\neq k}^M F_m\left(x\right)$ being the PDF of $\hat{S}$.
Obviously, the main idea behind the ML based ARSSS checking method is that the UE chooses the SS which is most likely to be a realization of $\hat{S}$.

Under the ML based ARSSS checking strategy, the SCR can be written as
\begin{align}
\mathcal{P}_S &= \mathcal{P}\left\{  f_{\mathrm{max}}\left(S_{M+1}\right) > \max_{1\leq m\leq M} f_{\mathrm{max}}\left(S_m\right)  \right\}
\nonumber \\
&= \int_{-\infty}^{+\infty}f_{M+1}(x)\prod_{m=1}^M \left\{\int_{\Omega\left(x\right)} f_{m}\left(t\right) \mathrm{d}t\right\}
\mathrm{d}x \label{MLCARSSSc}
\end{align}
where $\Omega\left(x\right)\triangleq\left\{t|f_{\mathrm{max}}(t)<f_{\mathrm{max}}(x)\right\}$ .

In general, numerical calculation of \eqref{MLCARSSSc} is involved due to the fact that $\Omega\left(x\right)$ does not have a closed form expression. For a special case when the UE is much closer to one of the LBS, for example LBS 1, we have $f_{\mathrm{max}}(x) \approx f_1(x)\doteq \frac{1}{\sqrt{2\pi}\sigma_S}\exp\left(-\frac{(x-u_1)^2}{2\sigma_S^2}\right)$, and thus
\begin{align}
\Omega\left(x\right)\approx\left\{
\begin{aligned}
\left\{t|t<x~\mathrm{or}~t>2u_1-x\right\},\quad x<u_1\\
\left\{t|t<2u_1-x~\mathrm{or}~t>x\right\},\quad x\geq u_1
\end{aligned}
\right..
\end{align}
\subsection{Numerical example \& Discussion}
\begin{figure}[t]
  \centering
  \includegraphics[width=3 in]{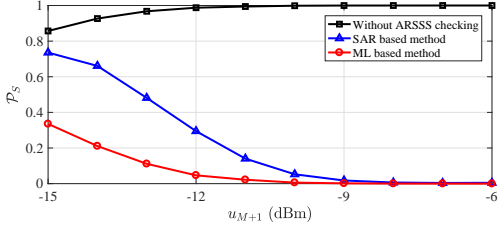}\\
  \caption{\small $\mathcal{P}_S$ versus $u_{M+1}$, where we set $M=3$, $d_1=80$ (m), $d_2=250$ (m), $d_3=250$ (m), $P=40~$(dBm), $\sigma_h^2=1$, $\sigma_\Psi^2=3$, $L = 10$, $\alpha = 3$, and $\delta=0.01$.}\label{PAC}
  \vspace{-5mm}
\end{figure}

In Fig. \ref{PAC}, we plot $\mathcal{P}_S$ versus $u_{M+1}$ when there is no ARSSS checking and when the SAR or ML based schemes are utilized.
As we can see, with the increase of $u_{M+1}$, the UE will connect to the FBS with probability approaching one.
However, with the proposed two ARSSS checking methods, for large $u_{M+1}$, the FBS will be easily distinguished by the UE, and in this cases, $\mathcal{P}_S$ becomes very small.
We note that each of the proposed schemes has its own advantages and disadvantages.
We observe that within a vast range value of $u_{M+1}$, the ML scheme outperforms the SAR scheme in term of suppressing the SCR $\mathcal{P}_S$. This is because the ML based scheme tends to find the SS that is most likely to be from the LBSs.
However, the complexity of the ML scheme is much higher than the SAR scheme. The SAR scheme only needs to compare the largest ARSSS with a pre-given number while the ML requires to calculate $f_{\max}(x)$, which involves exponential and Q functions, for $M+1$ times.
As for the SAR scheme, it associates the UE with the BS who provides the strongest signal strength out of the suspicious power region. As a result, conditioning on the UE will connect to a LBS, the BS selected by the SAR scheme may provide a higher link capacity than that by the ML scheme.

\section{Cooperation aided ARSSS checking}
\label{Sec:PLAScheme}
In this section, we introduce a method to further improve the performance of identifying the FBS by utilizing the cooperative nodes (CNs).
We assume there are several geometrically distributed and friendly cooperative nodes (CNs) around the UE \footnote{For example, the CNs can be other legitimate but idle UEs}. For simplicity, we only consider the cases where the UE is close to one LBS, and the ARSSSs from other LBSs is much lower than the closest one, and thus those LBSs  are neglected. More general scenarios are left for future works.
To facilitate the proposed method, reliable communication links between the UE and the CNs are required. This can be realized by the techniques such as device-to-device communication.

The basic idea here is that except for the UE, the CNs also record the ARSSSs of the their received SSs.
Then, the CNs feed their ARSSSs back to the UE. With these extra information provided by the CNs, the UE can make a more secure decision.

Denote  the PDF of the ARSSS from the LBS to the $i$-th CN as $f_{C,i}(x)$.
According to the central-limit theorem, we have $f_{C,i}(x)\doteq \frac{1}{\sqrt{2\pi}\sigma_S}\exp\left({-\frac{\left(x-u_{C,i}\right)^2}{2\sigma_S^2}}\right)$, where $u_{C,i}$ is the mean value of the ARSSS from the LBS to the $i$-th CN which can be obtained from the location information of the LBS and the $i$-th CN.
During the CS stage, both the UE and the CNs receives two strong SSs. Without loss of generality, the two SSs are denoted by $\bm{z}_1$ and $\bm{z}_2$. For $j\in\{1,2\}$, the ARSSSs of $\bm{z}_j$ at the UE and the $i$-th CN are denoted by $S_{j}$ and $S_{j,i}$, respectively.

With the help of the CNs, we summarize a detailed steps for the UE to distinguish the SS of the LBS as follows:
1) when the UE is much closer to one LBS but observes two strong SSs, i.e., $\bm{z}_1$ and $\bm{z}_2$, the UE broadcasts the index of the two indistinguishable SSs to nearby CNs;
2) The $i$-th CN observes the ARSSSs of these two SSs, and feeds back the values of $f_{C,i}\left(S_{1,i}\right)$ and $f_{C,i}\left(S_{2,i}\right)$;
3) The UE obtains the feedback from the CNs and determines that $\bm{z}_{j^*}$ is the SS from the LBS, where $j^*=\mathrm{argmax}_{j\in\{1,2\}}~f_1\left(Z_{j}\right)\prod_{i}f_{C,i}\left(S_{j,i}\right)$.

\begin{figure}[t]
  \centering
  \includegraphics[width=3 in]{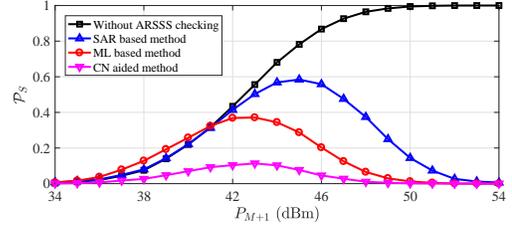}\\
  \caption{\small  $\mathcal{P}_S$ versus $P_{M+1}$, $P=40~$(dBm), $\sigma_h^2=1$, $\sigma_\Psi^2=3$, $L = 10$, $\alpha = 3$, $R_L=100~$m, $r_C=50~$m, $r_O=150~$m, $r_I=90~$m and $\delta=0.01$.}\label{PACR}
  \vspace{-5mm}
\end{figure}

In Fig. \ref{PACR}, we evaluate the performance of the proposed CN aided ARSSS checking method in terms of SCR. In our simulation, the UE and the LBS are located at $(0,0)$ and $(0,R_L)$, respectively. The simulation is carried out for $10000$ realizations. For each realization, two CNs are uniformly and randomly generated within $\mathcal{R}\left(r_C\right)$, and a FBS is uniformly and randomly generated within $\mathcal{R}\left(r_O\right)\setminus\mathcal{R}\left(r_I\right)$, where $\mathcal{R}\left(x\right)$ denotes the ball region whose center and radius are $(0,0)$ and $x$, respectively. As we can see from Fig. \ref{PACR}, with the extra information provided by the CNs, the SCR is greatly reduced.
This is mainly due to the fact that with more distributed nodes recording the ARSSS, we obtain a higher resolution when distinguishing the location of the source of the received SS, and with the prior location information of the LBSs, we can identify the FBS in a more reliable manner.
From Fig. \ref{PACR}, we can also observe that different from the cases without ARSSS checking, when the proposed methods are adopted, the SCR reaches the maximum when the FBS uses a moderate transmit power level. This is because in this case, the ARSSSs from the FBS and the LBS are less different, which makes it harder to distinguish between them.

\section{Conclusion}
\label{Sec:Conclusion}
In this letter, we proposed two location based schemes to combat the SS spoofing attack, i.e., the SAR and the ML based method.
We showed via numerical results that both the SAR and the ML based method deal well with the SS spoofing attack when the malicious FBS is greedy who adopts a large transmit power.
Besides, we also proposed a cooperation based method to further enhance the performance of combating the SS spoofing attack. It is shown that with the cooperation based method, the SDR can be greatly reduced in a vast range of spoofing power.

\end{document}